\documentclass[pra, a4paper, eqsecnum, preprint, showpacs,superscriptaddress]{revtex4-1}
\usepackage{amsmath,amscd,amsfonts,amssymb, amsbsy, stackrel, color,bbm, bm}
\usepackage{graphicx}
\usepackage{longtable}
\newcounter{matriz}
\newenvironment{matriz}{\refstepcounter{matriz}\equation}{\tag{A.\thematriz}\endequation}
\begin{document}
\title{Asymptotic spectral stability  of \\  the  Gisin-Percival state diffusion}  
\author{K. R. Parthasarathy}\email{krp@isid.ac.in}
\affiliation{Indian Statistical Institute, Theoretical Statistics and Mathematics Unit,Delhi Centre,
7 S.~J.~S. Sansanwal Marg, New Delhi 110 016, India} 
\author{A. R. Usha Devi}\email{arutth@rediffmail.com} 
\affiliation{Department of Physics, Bangalore University, 
Bangalore-560 056, India}
\affiliation{Inspire Institute Inc., Alexandria, Virginia, 22303, USA.}
\date{\today}
\begin{abstract} 
Starting from the  Gisin-Percival state diffusion equation for the pure state trajectory of a composite bipartite quantum system and exploiting the purification of a mixed state via its Schmidt decomposition, we write the diffusion equation for the quantum trajectory of the  mixed state of a subsystem $S$ of the bipartite system,  when the initial state in $S$ is mixed. Denoting the diffused state of the system  $S$ at time $t$ by  $\rho_t(\mathbf{B})$  for each $t\geq 0$, where $\mathbf{B}$ is the underlying  complex $n$-dimensional vector-valued Brownian motion process and using It{\^o} calculus, along with an induction procedure, we arrive at the stochastic differential of the scalar-valued moment process ${\rm Tr}[\rho_t^m( \mathbf{B})], \,\,\, m=2,3,\ldots$ in terms of $d\,\mathbf{B}$ and $d\,t$. This shows that each of the processes $\{{\rm Tr}[\rho_t^m( \mathbf{B})], t\geq 0\}$ admits a Doob-Meyer decomposition as the sum of a martingale $M^{(m)}_t(\mathbf{B})$ and a non-negative increasing process $S^{(m)}_t(\mathbf{B})$. This ensures the existence of 
$\underset{t\rightarrow\infty}{\lim}\, {\rm Tr}[\rho_t^m( \mathbf{B})]$ almost surely with respect to the Wiener probability measure $\mu$ of the Brownian motion $\mathbf{B}$, for each $m=2,\, 3,\, \ldots$.  In particular, when $S$ is a finite level system, the spectrum and therefore the entropy of $\rho_t (\mathbf{B})$ converge almost surely to a limit as $t\rightarrow \infty$.  In the Appendix, by employing probabilistic means, we prove a technical result which implies the almost sure convergence of the spectrum for countably infinite level systems.
\end{abstract}

\maketitle 
\section{Introduction} 
Let $\mathcal{H}$ be a complex Hilbert space describing the states of a quantum system. We consider the Gisin-Percival continuous time quantum diffusion trajectories~\cite{KRPARU, Gisin} $\{\vert\Psi_t(\mathbf{B})\rangle, t\geq 0\}$ with values on the unit sphere of the Hilbert space  $\mathcal{H}$, driven by a standard $n$-dimensional complex vector-valued  Brownian motion 
$\{\mathbf{B}(t)=(B_{1,t},B_{2,t},\ldots , B_{n,t}), t\geq 0\}$  with Wiener probability measure $\mu$ on the space of continuous paths: 
\begin{eqnarray} \label{dPsiPG}
 d\vert\Psi_t\rangle &=& \sum_{k=1}^n\, \widetilde{\mathbbm{L}}_{k,t}\, \vert\Psi_t\rangle\, \  dB_{k} 
-\left(i\, \widetilde{\mathbbm{H}}_t + \sum_{k=1}^{n}\,  \widetilde{\mathbbm{L}}_{k,t}^\dag\, \widetilde{\mathbbm{L}}_{k,t}
\right)\, \vert \Psi_t\rangle\, dt,   
\end{eqnarray}
where $\vert\Psi_0\rangle=\vert\phi_0 \rangle\in \mathcal{H}$ is the initial state. Here, we have denoted 
\begin{eqnarray} \label{TildeLH}
\widetilde{\mathbb{L}}_{k,t}&=&\mathbbm{L}_k-\langle \mathbbm{L}_k \rangle_{\Psi_t},\, \ \ 
\langle \mathbbm{L}_k \rangle_{\Psi_t}=\langle \Psi_t\vert\, \mathbbm{L}_k\,\vert \Psi_t\rangle  \nonumber \\ 
 \widetilde{\mathbbm{H}}_t&=& \mathbbm{H}+i\left(\mathbbm{L}_k\, \langle \mathbbm{L}^\dag_k\rangle_{\Psi_t}-\mathbbm{L}^\dag_k\, 
\langle \mathbbm{L}_k\rangle_{\Psi_t} \right),  
\end{eqnarray}
where $\mathbbm{L}_k, k=1,2,\cdots, n$ and $\mathbbm{H}$ are bounded operators in  $\mathcal{H}$, with $\mathbbm{H}$ being self-adjoint. 
  
We shall denote  the Hilbert space of $\mathcal{H}$-valued  norm square integrable functions \break $\{\vert\Psi_t(\mathbf{B})\rangle, t~\geq~0\}$ by $L^2(\mu)\otimes \mathcal{H}= L^2(\mu, \mathcal{H})$. 
The map $t\rightarrow \vert\Psi_t(\mathbf{B})\rangle$ is a non-anticipating state-valued Brownian functional in 
$L^2(\mu, \mathcal{H})$.   Our best estimate of all observable properties of the quantum system at a time instant $t\geq 0$ is reflected by the knowledge of the state diffusion trajectory up to that time. This, in turn, can be used  to predict the behavior of the system at a later time. Note that a pure state remains pure under the Gisin-Percival quantum state diffusion (\ref{dPsiPG}) at all times $t\geq 0$. However, a clear description on the nature of the {\em spectrum} of  a diffusion trajectory $\{\rho_t(\mathbf{B}),t \geq 0\}$ of mixed quantum states at a later time, based on the knowledge of such continuous time diffusion up to time $t$, demands a thorough analysis. Massen and K{\"u}mmerer~\cite{Massen} had investigated this topic in the case of a discrete time trajectory associated with a random chain of quantum states resulting from repeated measurements on a quantum system. Motivated by this work we study here the continuous time trajectory of a quantum system and arrive at a trace formula for the  scalar-valued moment processes $\{ {\rm Tr}[\rho^m_t(\mathbf{B})],\, t \geq 0\}$, $m=1,2,\ldots $ of  mixed states $\rho_t(\cdot)$ undergoing Gisin-Percival state diffusion. We show that $\{ {\rm Tr}[\rho_t^m(\mathbf{B})],t \geq 0\}$ for $m>1$ are submartingale processes~\cite{Doob, Williams}  in the space of continuous paths, with Wiener probability measure 
$\mu$. By the submartingale convergence theorem~\cite{Doob, Williams} it follows that  $\underset{t\rightarrow\infty}{\lim}\, {\rm Tr}[\rho^m_t(\mathbf{B})]$ exists almost surely for each of the bounded, non-negative, scalar-valued submartingale moment processes $\{0\leq {\rm Tr}[\rho^m_t(\mathbf{B})\leq 1, t\geq 0]\},\ m=2, 3,\, \ldots $.  
Thus, the spectrum of $\rho_t(\mathbf{B})$ converges almost surely with respect to the Wiener probability measure $\mu$  as $t\rightarrow \infty$. 

\section{Continuous time quantum diffusion trajectory of  mixed states}

Let us consider the Gisin-Percival equation (\ref{dPsiPG}) describing diffusion of pure states $\{\vert\Psi_t\rangle, t\geq0\}$ in 
$L^2(\mu, \mathcal{H})$, 
$\mathcal{H}=\mathcal{H}_{S}\otimes \mathcal{H}_{S'}$,
where  $\mathcal{H}_{S}$ and $\mathcal{H}_{S'}$ denote Hilbert spaces with  ${\rm dim}\,{\mathcal{H}_S}\geq {\rm dim}\,{\mathcal{H}_{S'}}$. We restrict to operator parameters $\mathbbm{L}_{k}=L_k\otimes I_{S'},$\ $\mathbbm{H}~=~H\otimes I_{S'}$, where $L_k,\ H\,$ are operators  in $\mathcal{H}_{S}$ and $I_{S'}$ is the identity operator in $\mathcal{H}_{S'}$. Thus,     
 \begin{eqnarray} \label{TildeLHS}
\widetilde{\mathbbm{L}}_{k,t}&=& \left( L_k -\langle L_k \otimes I_{S'} \rangle_{\Psi_t}\right)\, \otimes I_{S'},   \nonumber \\
                             &=& \widetilde{L}_{k,t}\otimes I_{S'},\ \ \ \ k=1,2,\ldots, n, \nonumber \\     
 \widetilde{\mathbbm{H}}_t&=& \left[\, H  +i\left(L_k\, \langle L^\dag_k\otimes I_{S'}\rangle_{\Psi_t}-
L^\dag_k\, \langle L_k\otimes I_{S'}\rangle_{\Psi_t} \right)\right]\otimes I_{S'}, \nonumber \\ 
                          &=& \widetilde{H}_t \otimes I_{S'}.    
\end{eqnarray}
Starting from a non-product and therefore, an entangled bipartite pure state $\vert\Psi_0\rangle\in \mathcal{H}_{S}\otimes\mathcal{H}_{S'}$, the Gisin-Percival state diffusion (\ref{dPsiPG}) results in a pure state quantum trajectory $\{\vert\Psi_t(\mathbf{B})\rangle, t\geq 0\}$, which is a non-anticipating Brownian functional with values on the unit sphere of $\mathcal{H}
~=~\mathcal{H}_{S}~\otimes~\mathcal{H}_{S'}$. We express  $\vert\Psi_t\rangle$  in terms of its Schmidt decomposition,     
\begin{equation} 
\vert\Psi_t\rangle=\sum_{\alpha}\, \sqrt{p_{\alpha,t}}\, \vert \alpha_{S}\otimes \alpha_{S'}\rangle_t, \ \  \sum_{\alpha}\, p_{\alpha,t}= 1,\ \ 
p_{\alpha, t}\geq 0 \,\ \  \forall\ \  t\geq 0,
\end{equation}
where $\{\vert \alpha\rangle_{S,\,t}\}$ and  $\{\vert \alpha\rangle_{S',\, t}\}$, are the set of eigenstates of the subsystem density matrices 
\begin{eqnarray*}
\rho_{S,t}&=&{\rm Tr}_{S'}[\,\vert\Psi_t\rangle\langle \Psi_t\vert\,]=\sum_\alpha\, p_{\alpha, t}\, \vert\alpha_S\rangle_t\langle\alpha_S\vert, \nonumber \\ 
 \rho_{S',t}&=&{\rm Tr}_{S}[\,\vert\Psi_t\rangle\langle \Psi_t\vert\,]=\sum_\alpha\, p_{\alpha, t}\, \vert\alpha_{S'}\rangle_t\langle\alpha_{S'}\vert.
\end{eqnarray*} 
 The eigenvalues (Schmidt coefficients) $p_{\alpha, t}\geq 0$ of the density matrices $\rho_{S,t}$, $\rho_{S',t}$ are arranged in the decreasing order $p_{1,t}\geq p_{2,t}\geq \ldots$. 

In this case, the operators $\widetilde{L}_{k,t}, \widetilde{H}_t$ of (\ref{TildeLHS}) take the form,     
 \begin{eqnarray} \label{TildeLHS2}
\widetilde{L}_{k,t} &=&  L_k -\langle  L_k\rangle_{t}  \nonumber \\                    
 \widetilde{H}_t&=& H  +i\left(L_k\, \langle L^\dag_k\rangle_{t}-
L^\dag_k\, \langle L_k\rangle_{t} \right)      
\end{eqnarray}
where, 
\begin{eqnarray}\label{TildeLHS3}
\langle  L_k\rangle_{t}&=&\langle L_k\otimes I_{S'} \rangle_{\Psi_t} = {\rm Tr}[\rho_{S,t}\, L_k].  
\end{eqnarray} 
Hereafter, our discussions will be centered on the properties of the quantum diffusion trajectory of  mixed states  $\{\rho_{S,t}(\mathbf{B}), t\geq 0\}$ in the space of density operators in $\mathcal{H}_S$ and hence, we shall write  
$\rho_{S,t} = \rho_{t}$, by dropping the suffix $S$ for brevity. 

\noindent{\bf Proposition}: Consider the state-valued  process 
$\{\vert\Psi_t(\mathbf{B}\rangle, t\geq 0\}$ on the unit sphere of $\mathcal{H}~=~\mathcal{H}_S\otimes \mathcal{H}_{S'}$ obeying  the  Gisin-Percival state diffusion equation (\ref{dPsiPG}), with $\widetilde{\mathbbm{L}}_{k,t}, k=1,2,\ldots, n,$ and \ $\widetilde{\mathbbm{H}}_t$ as in (\ref{TildeLHS}), (\ref{TildeLHS2}), and (\ref{TildeLHS3}). Then, $\{\rho_t(\mathbf{B})={\rm Tr}_{S'}[\,\vert 
\Psi_t(\mathbf{B})\rangle\langle\Psi_t(\mathbf{B})\vert\,], t\geq 0\}$ satisfies the following classical stochastic differential equation:
\begin{eqnarray} \label{rhot}
d\rho_t &=& \sum_{k=1}^{n}\,\left(\,\widetilde{L}_{k,t}\,\, \rho_t\, dB_{k}\, + \, \rho_t\,\widetilde{L}^\dag_{k,t}\, dB^*_{k}\,\right) \nonumber \\ 
&& \ \ \  +  \left\{ [\rho_t, i\, \widetilde{H}_{t}]
- \sum_{k=1}^{n}\, \left( \rho_t\, \widetilde{L}^\dag_{k,t}\, \widetilde{L}_{k,t} + 
\widetilde{L}^{\dag}_{k,t}\, \widetilde{L}_{k,t}\, \rho_t - 2\, \widetilde{L}_{k,t}\, \rho_t\,\widetilde{L}^\dag_{k,t}\,\right)\,\right\}\, dt.  
\end{eqnarray}

\noindent{\bf Proof:} Consider the Gisin-Percival state diffusion equation (\ref{dPsiPG}) in $L^2(\mu, \mathcal{H}_S\otimes\mathcal{H}_{S'})$, with an initial entangled bipartite pure state 
$\vert\,\Psi_0\rangle\in \mathcal{H}_S\otimes\mathcal{H}_{S'}$ and with the operator parameters  $\widetilde{L}_{k,t},\  \widetilde{H}_t$ of 
(\ref{dPsiPG}) as given in (\ref{TildeLHS2}), and (\ref{TildeLHS3}). Using the classical It{\^o} multiplication rule, 
\begin{equation}\label{itoB}
dB_{k,t}\,\, dB_{l,t}=0,\ dB_{k,t}\,\, dB^*_{l,t}=2\, \delta_{k\, l}\, dt,\ (dt)^2=0 
\end{equation}  
and simplifying, we obtain the following stochastic differential equation for the process $\{\vert\Psi_t(\mathbf{B})\rangle\langle\, \Psi_t(\mathbf{B})\vert, t\geq 0\}$:  
\begin{eqnarray}\label{projector}
&&d\, \left(\vert\Psi_t\rangle\langle\, \Psi_t\vert\right)= \left( d\, \vert\Psi_t\rangle\right)\, \langle\,\Psi_t\vert+ 
 \vert\Psi_t\rangle\, \left(d\langle\, \Psi_t\vert\right) + \left( d\, \vert\Psi_t\rangle\right)\,  \left(d\langle\, \Psi_t\vert\right)\nonumber \\
&& \ \ \ \  = \sum_{k=1}^{n}\,\left[\widetilde{L}_{k,t}\otimes I_{S'}\,\, \vert\Psi_t\rangle\langle\, \Psi_t\vert\, dB^*_{k}\, + 
\, \vert\Psi_t\rangle\langle \Psi_t\vert\,\,\widetilde{L}^\dag_{k,t}\otimes I_{S'}\,  dB_{k}\,\right] \nonumber \\ 
&& \ \ \ \ \ \ \ \  + \left\{\,  \left[\,\vert\Psi_t\rangle\langle\, \Psi_t\vert, i\, \widetilde{H}_{t}\otimes I_2\, \right]  - \sum_{k=1}^{n}\, 
\left( \vert\Psi_t\rangle\langle\, \Psi_t\vert\, \widetilde{L}^\dag_{k,t} \widetilde{L}_{k,t}\otimes I_{S'} \right. \right.\nonumber \\ 
&&\ \ \ \ \ \ \ \ \left. \left. +
\widetilde{L}^{\dag}_{k,t} \widetilde{L}_{k,t}\otimes I_{S'}\, \vert\Psi_t\rangle\langle\, \Psi_t\vert - 
2\, \widetilde{L}_{k,t}\otimes I_{S'}\, \vert\Psi_t\rangle\langle\, \Psi_t\vert\, \widetilde{L}^\dag_{k,t}\otimes I_{S'}\,\right)\right\}\, dt. 
\end{eqnarray}
Taking partial trace  over $S'$ in (\ref{projector}) results in (\ref{rhot}). \hskip 2in $\square$

\noindent {\em Remark:} Since $\widetilde{L}_{k,t}$, $\widetilde{H}_t$ are related to $L_k$, $H$ (see (\ref{TildeLHS2})) by {\em translation}  via scalar quantities   ${\rm Tr}\, [\rho_t\, L_k]$, $k=1,2,\ldots , n$,   the stochastic differential equation (\ref{rhot}) can be rewritten as  (see Section IV of Ref.~\cite{KRPARU} for a discussion on the {\em translational invariance} of the Gorini-Kossakawski-Sudarshan-Lindblad (GKSL) generator~\cite{GKS, Lindblad}) of the quantum dynamical semigroup $\{T_t, t\geq 0\}$) 
\begin{eqnarray} \label{rhotinv}
d\rho_t &=& \sum_{k=1}^{n}\,\left( \widetilde{L}_{k,t}\,\, \rho_t\, dB_{k}\, + \, \rho_t\,\widetilde{L}^\dag_{k,t}\, dB^*_{k}\,\right) \nonumber \\ 
&&\ +  \left( [\rho_t, i\, H]
- \sum_{k=1}^{n}\, \left( \rho_t\, L^\dag_{k}\, L_{k} + 
L^{\dag}_k\, L_{k}\, \rho_t - 2\, L_{k}\, \rho_t\,L^\dag_{k}\,\right)\,\right)\, dt,  
\end{eqnarray}
 it follows that (i) $\{\rho_t(\mathbf{B}),t\geq 0\}$  obeys a diffusion equation; (ii) it is a Markov process with initial value 
$\rho_0$ and with the infinitesimal generator $\mathcal{L}$ at $\rho$  given by  
\begin{eqnarray} \label{infgen}
\mathcal{L}(\rho)&=&    [\rho, i\, H]
- \sum_{k=1}^{n}\, \left( \rho\, L^\dag_{k}\, L_{k} + 
L^{\dag}_{k}\, L_{k}\, \rho - 2\, L_{k}\, \rho\,L^\dag_{k}\,\right),  
\end{eqnarray}
  in the  GKSL form~\cite{GKS, Lindblad}.

 From the stochastic differential equation (\ref{rhot}), and equivalently (\ref{rhotinv}), for the  quantum trajectory $\{\rho_t(\mathbf{B}), t\geq 0\}$ it follows immediately that  $d\, {\rm Tr}[\,\rho_t(\mathbf{B})\,]=0$, with intial condition 
${\rm Tr}[\,\rho_0\,]=\langle \Psi_0\,\vert \Psi_0\rangle=1$. Thus, ${\rm Tr}[\,\rho_t (\mathbf{B})\,]=1$ for all $t\geq 0$. In other words, (\ref{rhot}) and (\ref{rhotinv}) are, indeed, state diffusion equations.

\section{The scalar-valued  moment processes  $\{{\rm Tr}[\rho_t^m(\mathbf{B})], t\geq 0\}$ and asymptotic spectral stability} 

Based on the Gisin-Percival diffusion equation (\ref{rhot})  we now present the following Theorem. 

\noindent{\bf Theorem 1:} The  processes $\{\rho_t^m(\mathbf{B}), t\geq 0\}, m=2,3,\ldots$ satisfy the stochastic differential equations   
\begin{eqnarray} \label{rhotm} 
  d\, \rho_t^m &=& 2\, \sum_{k=1}^{n}\,  {\rm Re}\, \left( \sum_{r=0}^{m-1}\, \rho_t^r\, \widetilde{L}_{k,t}\, \rho^{m-r}_t\ dB_{k}\right)    \nonumber \\
&& \ \ \ \  +\left\{  \widetilde{\mathcal{L}}(\rho_t^m) + 2\, \sum_{k=1}^n\, \left(\mathop{\sum_{m_1+m_3, m_2\neq 0,}^{}}_{m_1+m_2+m_3=m}\,    \rho_t^{m_1}\, \widetilde{L}_{k,t}\, \rho_t^{m_2}\, \widetilde{L}_{k,t}^\dag\, \rho_t^{m_3} \right. \right. \nonumber \\ 
&&\ \ \ \ \left.\left.    +   \, \mathop{\sum_{m_1,\, m_2\neq 0}^{}}_{m_1+m_2+m_3=m}\,  \rho_t^{m_1}\, \widetilde{L}^\dag_{k,t}\, \rho_t^{m_2}\, \widetilde{L}_{k,t}\, \rho_t^{m_3}\, \right) \right\} dt   
\end{eqnarray}
where  
\begin{eqnarray} \label{lindblad}
\widetilde{\mathcal{L}}(\rho_t^m)&=& \left[\,\rho_t^m,\, i\, \widetilde{H}_t\, \right]- \sum_{k=1}^n\,\left(\rho_t^m\, \widetilde{L}_{k,t}^\dag\,\widetilde{L}_{k,t} + \widetilde{L}^\dag_{k,t}\,\widetilde{L}_{k,t}\, \rho_t^m - 
2\, \widetilde{L}_{k,t}\, \rho_t^m\, \widetilde{L}_{k,t}^\dag\right).  
\end{eqnarray}
The summation in the second and third lines of (\ref{rhotm}) involves positive integers $m_1,m_2,m_3$ such that $m_1+m_2+m_3=m$.   
             
\noindent{\bf Proof:} We derive the stochastic differential equation satisfied by $\rho_t^2(\mathbf{B})$  using (\ref{rhot}) and by simple application of It\^o's classical stochastic calculus~\cite{Mckean}: 
 \begin{eqnarray}\label{rho2}
 d\, \rho_t^2 &=& (d\, \rho_t)\, \rho_t +\, \rho_t\, (d\, \rho_t)+ (d\, \rho_t)^2\nonumber \\ 
 &=& 2\, \sum_{k=1}^{n}\,  {\rm Re}\, \left[\left(\widetilde{L}_{k,t}\, \rho^{2}_t + \rho_t\, \widetilde{L}_{k,t}\, \rho_t\,\right)\, 
d\, B_{k}\, \right] \nonumber \\
&&\ \ \ \    + \left\{ \widetilde{\mathcal{L}}(\rho_t^2) + 2\, \sum_{k=1}^{n}\,\left(\, \rho_t \widetilde{L}_{k,t}\, \rho_t\, \widetilde{L}^\dag_{k,t} 
+ \,   \widetilde{L}_{k,t}\,\rho_t 
\widetilde{L}^\dag_{k,t}\, \rho_t  \, \right) \, \right\}\  dt,   
\end{eqnarray}
which is in agreement with (\ref{rhotm}) for $m=2$. Then, it immediately follows by mathematical induction that if (\ref{rhotm}) holds for some positive integer $m$, it also holds for $m+1$. \ \hskip 0.2in $\square$ 

We now state our result on the scalar-valued moment processes ${\rm Tr}[\, \rho_t^m(\mathbf{B})], m=2,3,\ldots$  of the continuous time quantum diffusion trajectory $\rho_t(\cdot)$.

\noindent {\bf Theorem 2:} Under the Gisin-Percival continuous time diffusion (\ref{rhot}) the non-negative bounded scalar-valued  moment processes  $0\leq {\rm Tr}[\, \rho_t^m(\mathbf{B})]\leq 1,\ \  m=2,3,\ldots$ of the quantum trajectory $\{\rho_t(\mathbf{B}),\, t\geq 0\}$ admit the following stochastic differentials:
\begin{eqnarray} \label{moment}
d\, {\rm Tr}\,[\, \rho_t^m]&=&2\, m\, \sum_{k=1}^n\, {\rm Re}\,\left( {\rm Tr}[\rho_t^m\, \widetilde{L}_{k,t}]\, dB_{k}\right)  + 
2\, m\, \sum_{k=1}^n\left(\sum_{m'=1}^{m-1}\, {\rm Tr}[\rho_t^{m'}\, \widetilde{L}_{k,t}\, \rho_t^{m-m'}\, \widetilde{L}_{k,t}^\dag]\,\right)\, dt.  \nonumber \\
\end{eqnarray}

\noindent{\bf Proof:}  Taking trace in (\ref{rhotm}) and  noting   that  
${\rm Tr}[\,\widetilde{\mathcal{L}}\,(\,\cdot\,)\,]=0$ (see (\ref{lindblad})), one obtains the stochastic differential equations (\ref{moment}) for the scalar-valued moment processes ${\rm Tr}[\, \rho_t^m(\mathbf{B})],\ m=2,3,\ldots$. \hskip 3in $\square$  

\noindent{\bf Corollary 1:} The scalar-valued moment process ${\rm Tr}\,[\, \rho_t^m(\mathbf{B}), t\geq 0],$ admits the Doob-Meyer decomposition~\cite{Doob, Williams} 
\begin{equation} \label{dm}
 {\rm Tr}\,[\, \rho_t^m(\mathbf{B})]=\, M^{(m)}_t(\mathbf{B}) \, +\, \mathbf{S}^{(m)}_t(\mathbf{B}),
\end{equation}
where $\{M^{(m)}_t(\mathbf{B}), t\geq 0\}$  is the martingale given by 
\begin{eqnarray} 
M^{(m)}_t(\mathbf{B}) &=& {\rm Tr}[\rho_0^m\,]+\, 2\,m\, \int_{0}^{t}\,  \sum_{k=1}^n\, {\rm Re}\,\left({\rm Tr}[\rho_s^m\, \widetilde{L}_{k,s}]\, dB_{k,s}\right)
\end{eqnarray}
and $\{S^{(m)}_t(\mathbf{B}), t\geq 0\}$ is the non-negative increasing process given by 
\begin{eqnarray} \label{stb} 
S^{(m)}_t(\mathbf{B}) &=& 2\, m\, \int_{0}^{t}\,  \sum_{k=1}^n\, \left(\sum_{m'=1}^{m-1}\, {\rm Tr}[\rho_s^{m'}\, \widetilde{L}_{k,s}\, \rho_s^{m-m'}\, \widetilde{L}_{k,s}^\dag]\, ds\, \right)
\end{eqnarray}

\noindent{\bf Proof:} Immediate from Theorem 2 and the fact that each trace term on the right hand side of (\ref{stb}) is nonnegative. 
\hskip 4in $\square$ 

\noindent {\em Remark:} It follows from the Doob-Meyer decomposition (\ref{dm})  that the scalar-valued moments ${\rm Tr}\,[\, \rho_t^m(\mathbf{B}), t\geq 0]$  increase on average i.e., 
\begin{eqnarray}\label{submartingale}
\mathbbm{E}_s\left\{\, {\rm Tr}[\rho^m_t(\mathbf{B})]\, \vert \mathbf{B}(s),\ t\geq s\, \right\}\, \geq\, 
{\rm Tr}[\rho^m_s(\mathbf{B})]. 
\end{eqnarray}

\noindent{\bf Corollary 2:} For each $m=2,3,\ldots $ 
 $$\underset{t\rightarrow\infty}{\lim}\, {\rm Tr}\, [\,\rho_t^m(\mathbf{B})\,], \ \ \ {\rm a.s} \ \ \mathbf{B}\ (\mu)$$ 
exists with respect to the Wiener probability measure $\mu$. 

\noindent{\bf Proof:} From Corollary 1 it follows that $\{{\rm Tr}\,[\rho_t^m(\mathbf{B})]\}$ is a nonnegative bounded submartingale for each $m=2,3,\ldots$ and hence, the required convergence  is a consequence of the submartingale convergence theorem~\cite{Doob, Williams}. \hskip 2.5in $\square$

\noindent{\bf Corollary 3:} Equations (\ref{moment}) can be expressed in terms of the resolvent~\cite{Kato} $\left(1-x\, \rho_t\right)^{-1}$ of $\rho_t$, where $-1 < x < 1$, as follows: 
\begin{eqnarray}\label{resolvent} 
d\, {\rm Tr}\,[\left(1-x\, \rho_t\right)^{-1}]&=& 2\, \sum_{k=1}^{n}\, {\rm Re}\left(\frac{d}{d\,x}\left\{{\rm Tr}\left[\left(1-x\, 
\rho_t\right)^{-1}\, \rho_t\, L_{k,t}\right]\, d\, B_{k}\right\}\right) \nonumber \\ 
&& \ \ +2\, \sum_{k=1}^{n}\, \frac{d}{d\,x}\left\{{\rm Tr}\left[ \rho_t\, L_{k,t}\left(1-x\, 
\rho_t\right)^{-1}\, L_{k,t}^\dag\, \left(1-x\, 
\rho_t\right)^{-1}  \right]\, \right\}\, dt.
\end{eqnarray}     
\noindent {\bf Proof:} Immediate from the properties of the resolvent~\cite{Kato}. \hskip 1in $\square$

\noindent{\bf Corollary 4:} Let $S$ be a finite dimensional Hilbert space of dimension $d$.  Suppose $p_{1,t}(\mathbf{B})\geq p_{2,t}(\mathbf{B})\geq \ldots \geq p_{d,t}(\mathbf{B})$ is an enumeration of the eigenvalues of $\rho_t(\mathbf{B})$ in Theorem 2. Then, 
$$\underset{t\rightarrow\infty}{\lim}\, p_{\alpha,t}(\mathbf{B})\ \ {\rm a.s}\ \ \mathbf{B}\ \ (\mu)\ \ {\rm exists}$$ 
for every $1\leq \alpha\leq d$ with respect to the Wiener probability measure $\mu$. 

\noindent{\bf Proof:} This is immediate from Corollary 2, Theorem 2.
\hskip 1in $\square$ 

\noindent {\em Remark:} Corollary 4 implies that, when $S$ is a finite level quantum system, the Gisin-Percival state diffusion trajectory for any mixed initial state $\rho_0$ in $\mathcal{H}_S$ has an asymptotically stable spectrum  almost surely. In the Appendix we prove the almost sure convergence of the spectrum for countably infinite level systems using a probabilistic approach.
However, in the infinite dimensional case, the sum of the limits of all eigenvalues ${\displaystyle\sum_{\alpha}}\underset{t\rightarrow\infty}{\lim}\, p_{\alpha,t}$ can be strictly less than unity with a positive probability. In other words, the trajectory of the state diffusion in the infinite dimensional case can get knocked out of the set of  those described by  density operators. 

\section*{Appendix}
\noindent
Let the system Hilbert space $\mathcal{H}_S$ be equipped with a finite or countable infinite orthonormal basis and let 
$t\rightarrow \sigma_t$ be a map from the interval $[0,\infty)$ to the space of density operators in $\mathcal{H}_S$ such that for any fixed  $t$, the eigenvalues $p_{\alpha, t},\ \alpha=1,2,\ldots,$  of  
$\sigma_t$ are enumerated in decreasing order, inclusive of their multiplicity, as   
\begin{matriz} 
\label{a1}
p_{1,t}\geq p_{2,t}\geq \ldots \geq 0, 
\end{matriz} 
\begin{matriz} 
\label{a2}
\sum_{\alpha\geq 1}\, p_{\alpha, t}=1.
\end{matriz}     
We assume that the limits
\begin{matriz}
\label{a3}
\underset{t\rightarrow\infty}{\lim}\, {\rm Tr}\,[\sigma_t^m]=\underset{t\rightarrow\infty}{\lim}\, \sum_{\alpha\geq 1}\, p_{\alpha, t}^m=\kappa_m
\end{matriz} 
exist for each $m=1,2,\ldots,$ and, by definition,  $\kappa_1=1$.  Then the following theorem holds.

\noindent{\bf Theorem:} There exists a sequence $\{p_\alpha, \alpha\geq 1\}$ satisfying the following: 
\begin{matriz}
\label{a4}
p_1\geq p_2 \geq \ldots \geq 0,
\end{matriz}
\begin{matriz} 
\label{a5}
\sum_{\alpha\geq 1}\, p_{\alpha}\leq 1, 
\end{matriz}
\begin{matriz}
\label{a6}
 \underset{t\rightarrow\infty}{\lim}\,p_{\alpha,t}=p_\alpha,\ \alpha\geq 1.
\end{matriz}

\noindent {\bf Proof:} For each $0\leq t<\infty$, introduce a random variable $\xi_t$ assuming the values $p_{\alpha, t}$ with respective probabilities $p_{\alpha, t}, \alpha\geq 1,$ so that 
\begin{eqnarray*}
\mathbbm{E}\, \xi_t^m= \sum_{\alpha\geq 1}\, p_{\alpha, t}^{m+1} \hskip 1.5in 
\end{eqnarray*} 
\begin{matriz} 
\label{a7}
={\rm Tr}\,[\sigma_t^{m+1}],\ \ m=0,1,2,\ldots .    
\end{matriz}

Denote by $\mu_t,$ the probability measure, which is the distribution of  $\xi_t$. Each $\mu_t$  is a probability measure in the compact interval $[0,1]$. By equations (\ref{a3}) and (\ref{a7}) it follows that the $m^{\rm th}$ moment of the distribution $\mu_t$ converges to $\kappa_{m+1}$ for each $m$ as $t\rightarrow\infty$. Hence there exists a probability measure $\mu$ in the interval [0,1] such that $\mu_t$ converges {\em weakly}~\cite{Patrick} to $\mu$ as $t\rightarrow \infty$ i.e., for every real continuous function $\phi$ on [0,1],  
\begin{matriz}
\label{a8}
\underset{t\rightarrow\infty}{\lim}\, \int_0^1\, \phi(x)\, \mu_t(dx)
=\int_0^1\, \phi(x)\, \mu(dx).
\end{matriz}
(Indeed, this is a consequence of the fact that the space of all probability measures in the compact metric space [0,1] is a compact metric space in the topology of weak convergence and moments determine a distribution uniquely~\cite{Patrick, KRP_M}).

Now our goal is to determine the spectrum of $\mu$ i.e., the smallest closed set with $\mu$-probability equal to 1. To this end, we choose and fix a sequence 
\begin{matriz}
\label{a9}
0< t_1< t_2<\ldots 
\end{matriz}
by the diagonalization procedure, such that $t_n\rightarrow \infty$ as $n\rightarrow \infty$ and, 
\begin{matriz}
\label{a10}
\underset{n\rightarrow\infty}{\lim}\, p_{\alpha,t_n}=p_\alpha,
\end{matriz}
exists for every $\alpha\geq 1$. Then, 
\begin{matriz}
\label{a11}
p_1\geq p_2\geq \ldots \geq 0.
\end{matriz}
By Fatou's lemma~\cite{KRP_M}, 
\begin{matriz}
\label{a12}
1=\underset{n\rightarrow\infty}{\lim}\,\sum_{\alpha\geq 1} p_{\alpha, t_n} \geq \sum_{\alpha\geq 1}\, p_\alpha.
 \end{matriz}
Now three cases arise: 

\noindent{\bf Case (i):} $p_1=0$. 

By (\ref{a11}), it follows that  $p_2=p_3=\ldots =0$. By choosing $\phi(x)=x$ in (\ref{a8}), we get  
\begin{matriz}
\label{a13}
\underset{t\rightarrow\infty}{\lim}\, \int_0^1\, x\, \mu_t(dx)
=\int_0^1\, x\, \mu(dx).
\end{matriz}
Hence, 
\begin{matriz}
\label{a14}
\underset{n\rightarrow\infty}{\lim}\, \sum_{\alpha\geq 1}\, 
p^2_{\alpha,t_n}=\int_0^1\, x\, \mu(dx).
\end{matriz}
As $\sum_{\alpha\geq 1}\, p^2_{\alpha,t_n}\leq p_{1, t_n},$
we obtain 
$$\int_0^1\, x\, \mu(dx)=0,$$
which implies that $\mu$ is a probability measure degenerate at 0. 
Now (\ref{a14}) leads to   
$$\underset{t\rightarrow\infty}{\lim}\, \sum_{\alpha\geq 1}\, 
p^2_{\alpha,t}=0.$$ 
Thus, 
$$\underset{t\rightarrow\infty}{\lim}\,  
p_{\alpha,t}=0, \ \ \ \forall \ \ \ \alpha\geq 1.$$
 
\noindent{\bf Case (ii):} $p_1=1$. 

In this case, (\ref{a11}) and (\ref{a12}) imply
$$p_2=p_3=\cdots =0.$$
By (\ref{a14}), 
$$\underset{n\rightarrow\infty}{\lim}\, \sum_{\alpha\geq 1}\, 
p^2_{\alpha,t_n}\geq \underset{n\rightarrow\infty}{\lim}\, p^2_{1, t_n}=1.$$  
Thus, 
$$ \int_0^1\, x\, \mu(dx)=1.$$
This is possible only if $\mu$ is degenerate at the point 1. Thus, 
by (\ref{a14}) 
$$\underset{t\rightarrow\infty}{\lim}\, \sum_{\alpha\geq 1}\, 
p^2_{\alpha,t}=1.$$ 
Since  $\displaystyle\sum_{\alpha\geq 1}\, 
p_{\alpha,t}=1$,  it follows that 
$$\underset{t\rightarrow\infty}{\lim}\, p_{1,t}=1,\ \ 
\underset{t\rightarrow\infty}{\lim}\, p_{\alpha,t}=0, \ \ {\rm for} \ \ \ \alpha\geq 2.$$ 

Thus, the theorem needs to be proved only in Case (iii). 

\noindent{\bf Case (iii):} $0< p_1< 1.$ 

Now there exist $\alpha_1,\ \alpha_2,\ldots,\ $ and $1> q_1>q_2>\ldots >0$ such that   
\begin{eqnarray*}
&& p_1=p_2=\ldots =p_{\alpha_1}=q_1 \\ 
&& p_{\alpha_1+1}=p_{\alpha_1+2}=\ldots =p_{\alpha_1+\alpha_2}= q_2\, <\, q_1 \\ 
&& \hskip 1in  \vdots  \hskip 0.5in \vdots   \\  
&& p_{\alpha_1+\alpha_2+\ldots +\alpha_{r-1}+1}=p_{\alpha_1+\alpha_2+\ldots +\alpha_{r-1}+2}=\ldots = 
p_{\alpha_1+\alpha_2+\ldots +\alpha_{r}}
= q_r < q_{r-1} \\ 
&& \hskip 1in  \vdots  \hskip 0.5in \vdots   \\  
\end{eqnarray*}
which may be a terminating or a non-terminating sequence.  

Since $q_1=p_1$ and $0< q_1<1$, choose an arbitrary $\epsilon\geq 0$ such that $0<q_1+\epsilon<1$ and consider the open set $(q_1+\epsilon, 1]$ in the compact space $[0,1]$. Since, $p_{1, t_n}\rightarrow q_1$ as $n\rightarrow \infty$, we have 
$$  p_{1, t_n}\leq  q_1+\epsilon\ \ {\rm for \ all \ large}\ n$$
and therefore
$p_{\alpha,t_n}\leq q_1+\epsilon$ for all $\alpha\geq 1$ and for all large $n$. Thus $\mu_{t_n}$ has its support in $[0, q_1+\epsilon]$ for large $n$. Hence the support of $\mu$ is contained in $[0, q_1+\epsilon].$ Arbitrariness in $\epsilon$ implies that  the support of $\mu$ is contained in $[0,q_1]$. 

Now consider an open interval $(q_2+\epsilon, q_1-\epsilon)\subset 
[q_2,q_1]$, where $\epsilon$ is arbitrary, positive, but  $\epsilon <\frac{q_1-q_2}{2}$. Then, 
\begin{eqnarray*}
{\rm max}\, \left(p_{\alpha_1+1,t_n},\ p_{\alpha_1+2,t_n},\ \ldots 
p_{\alpha_1+\alpha_2,t_n}\right)\leq q_2+ \epsilon \\ 
{\rm min}\, \left(p_{1,t_n},p_{2,t_n},\ldots 
p_{\alpha_1,t_n}\right)\geq q_1 -\epsilon  
\end{eqnarray*}  
for all sufficiently large $n$. In other words, 
$$\mu_{t_n}\left((q_2+\epsilon, q_1-\epsilon)\right)=0$$
for all large $n$ and hence, 
$$\mu\left((q_2+\epsilon, q_1-\epsilon)\right)=0.$$
The arbitrariness in $\epsilon$ implies 
$$\mu\left((q_2, q_1)\right)=0.$$
By a similar argument we obtain 
$$\mu\left((q_{r+1}, q_r)\right)=0$$
whenever $q_{r+1}>0.$ Thus the spectrum of $\mu$ is contained in 
$\{q_1,q_2,\ldots , \} \cup \{0\}.$ 

By (\ref{a8}), for any continuous function $\phi$, 
\begin{matriz}
\label{a15}
 \underset{n\rightarrow\infty}{\lim}\, \sum_{\alpha\geq 1}\, \phi(p_{\alpha, t_n})\, p_{\alpha, t_n}=\, \sum_{r}\, \phi(q_r)\, 
\mu\left( \{\,q_r\,\} \right). 
\end{matriz} 
Choose $\phi$ to be the  function defined by    
\begin{matriz}
\label{a16} 
\phi(x)=\left\{ \begin{array}{l} 1,\  \ x\in [q_{s}-\epsilon, q_{s}+\epsilon] \\ 
0,\ {\rm if\ } x\notin \, (q_{s}-2\epsilon, q_{s}+2\epsilon)  \\ 
{\rm linear\ in}\ \   [q_{s}-2\epsilon, q_{s}-\epsilon]\,\,\cup\,\, 
[q_{s}+\epsilon, q_{s}+2\epsilon].    \end{array} \right.
\end{matriz} 
Then, (\ref{a15}) takes the form 
\begin{matriz}
 \underset{n\rightarrow\infty}{\lim}\, \sum_{j=1}^{\alpha_s}\, 
\phi(p_{\alpha_1+\alpha_2+\ldots +\alpha_{s-1}+j, t_n})\, p_{\alpha_1+\alpha_2+\ldots +\alpha_{s-1}+j, t_n}=
 \mu\left( \{\,q_s\,\} \right). 
\end{matriz}
or 
$$\alpha_s\, q_s=\mu\left( \{\,q_s\,\}\right)$$
for all $s>1$. The same holds for $s=1$ with a slight (and obvious) modification in the choice of $\phi$. 

Thus the limit $\{p_\alpha\}$  is independent of  the choice of the  diagonalization procedure. In other words, 
\begin{eqnarray*} 
\underset{t\rightarrow\infty}{\lim}\, p_{\alpha, t}=p_\alpha, \ 
\ \ \forall \ \ \ \alpha\geq 1.
\end{eqnarray*}
thus ensuring the convergence of the spectrum. \hskip 0.5in $\square$

\section*{Acknowledgements}
Major part of this work was done when the second author (ARU) was visiting Indian Statistical Institute, Delhi, during her sabbatical leave  from Bangalore University; her research is supported by the Major Research Project (Grant No. MRPMAJOR-PHYS-2013-29318) of the University Grants Commission (UGC), India.

\end{document}